# Laser-driven jetting of nanoscale non-conducting liquid droplets via hollow optical fiber

Jinwon Yoo,[1] Honggu Choi,[1] Om Krishna Suwal,[1] Sungrae Lee,[1] Woohyun Jung,[1] Sung Hyun Kim,[2] Sun-Mi Lee,[2] Kyung-hwa Yoo,[2] WonHyoung Ryu,[3] and Kyunghwan Oh[1]

[1]*Photonic Device Physics Laboratory, Department of Physics, Yonsei University, 50 Yonsei-ro Seodaemun-gu, Seoul 03722, Korea*
[2]*Nanodevice and Nanobio Laboratory, Departmenet of Physics, Yonsei University, 50 Yonsei-ro, Seoul 03722, Republic of Korea*
[3]*Biomedical and Energy System Laboratory, Department of Mechanical Engineering, Yonsei University, 50 Yonsei-ro, Seodaemun-gu, Seoul 03722, Korea*

**Abstract:** Along a single strand of micro-capillary optical waveguide, we achieved an efficient transfer of the light momentum onto the liquid contained there within, successfully atomizing it into nanoscale droplets. A hollow optical fiber (HOF), with a ring core and central air hole, was used to optically drive jetting of non-conducting transparent liquid of sub-pico liter volume, out of a surface-treated facet orifice, producing droplets ranging from nano to micrometer scale. These droplets were carried over the propagating light field forming a spherical cone, which were then deposited on a silica substrate in a Gaussian spatial distribution. The deposited patterns and sizes of individual droplets were characterized as a function of the laser power, irradiation time, and distance between the HOF and a substrate. This HOF based laser driven atomization technique obviates imperative electrode or aerial pressure requirements in prior methods, opening a new pathway to drastically scale down the form-factor of liquid jetting devices, and has a high potential to in-situ atomization and delivery of bio-medical non-conducting liquids in a microscopic environment, which was not possible in prior arts.

## 1. Introduction

Conventional microscopic jetting technology systematically generates liquid droplets or particles of micrometer size to propel them on various substrates and is being further developed for various emerging printing applications [1-3]. Among those technologies, electro-hydrodynamic (EHD) technique has been intensively explored in recent years for its high potential providing a precise spatial resolution, by applying strong electric fields to eject and direct charged liquid droplets [4-6]. Due to the requirement of sufficient amount of electrical charges in the liquid, conducting solutions are being heavily preferred in EHD printing. Although there have been attempts of using hydrophobic compounds with EHD jetting, additional processes such as removal of hydrophilic solvents or charge species have made this approach less attractive [7,8]. Despite some of impressive success in EHD, for a volume of liquid as little as sub-pico liter, there has been no appropriate technical solution to stable liquid atomization and delivery issue. Use of strong electrical fields poses additional potential risk of damaging fragile molecules in solutions, which has hindered broadening EHD applications further into bio-medical areas. Furthermore, imperative requirement of two electrodes in EHD has been a serious bottleneck in miniaturization for in-situ applications [9-11].

In this paper, we present an all-laser-driven liquid jetting technology that can eject "non-conducting" liquids (including hydrophobic and lipophilic) droplets through a single strand of hollow optical fiber (HOF) without external electric field or gas flow, for the first time to the best knowledge of the authors. This technique uses only irradiation of a continuous wave (CW) mid-infrared (IR) laser with a moderate power through a HOF filled with the liquid of interest to generate atomized liquid droplets then carry them over the light field propagating toward a solid substrate. Inherent flexibility of optical fiber provides versatile and accurate deployment of the proposed system, which enables in-situ jetting of sub-pico liter volume liquid in a microscopic environment, which was not possible in prior arts.

## 2. Results and Discussion

The schematic diagram of the proposed method is summarized in Fig. 1(a). HOF consisted of three layers: central air hole, germano-silicate high index ring core, and silica cladding [12], see the inset micro-photograph. The laser initially propagates along the ring core and then into the transparent liquid filled in the hole whose refractive index is higher than that of the ring core. This changes in the light propagation produces an efficient light momentum transfer to the liquid. At the end facet of the HOF, diamond nano-particles were deposited to provide an optimal surface energy for the liquid jetting. As the liquid was thrusted by the light momentum along HOF, droplets were formed at the end facet orifice, and subsequently they were carried over the light field coming out of the fiber to be deposited on a substrate. Note that liquid thrust, atomization of liquid droplets, and their transfer were all achieved by using only the laser light, which can provide a new avenue of liquid atomization technology.

The HOF was designed to guide the laser at the wavelength of 980nm in the ring core, which has the refractive index of $n_{core}$=1.457. HOF had the central air hole diameter of 3.1μm, surrounded by the ring core with the thickness of 2.6μm and cladding diameter of 125μm, which ensures adiabatic mode transformation between conventional single mode fibers (SMFs) as in other fiber optics applications [13,14].

In order to facilitate this light assisted momentum transfer, the liquid should have a high optical transparency and high dielectric constant at the laser wavelength. Firstly, we used glycerol as a non-conducting liquid, for its low absorption at the laser wavelength λ=980 nm [16], high dielectric constant of 41 [17], and refractive index of $n$=1.463 [18] higher than that of the HOF core. In fact, glycerol has been widely used to understand the mechanism of droplet formation in prior studies as it has a relatively high viscosity of 1200mPa·S at room temperature [19].

In this experiment, we found that successful jetting of glycerol from HOF required additional preparation such that the end facet of the HOF and its inside surface should be covered with a rough surface layer. It is well-known that the solid-liquid interaction at their interface can be significantly affected by introducing additional rough nano-scale layer [20], and we used diamond nano-particles to roughen the HOF silica surfaces at the exit facet. Diamond nano-particles with the diameter of ~10nm, were deposited at the end facet of the HOF, by dipping the HOF in a colloidal suspension solution similar to prior publications [21,22]. The prepared HOF is shown in Fig. 1(b) along with a glycerol filled to the length of ~300μm in the central hole. Glycerol was filled by the capillary force by immersing HOF in the glycerol reservoir for 10~30 seconds, and its volume was estimated as ~2.6 pico-liter. Fig. 1(c) shows the schematic diagram of the experimental setup. We spliced ~1.5 cm long HOF with single mode fiber (SMF), which was connected to a CW IR diode laser lasing at the wavelength of 980nm. The glycerol filled HOF was positioned vertically above a silica ($SiO_2$) substrate using a xyz translation stage. We launched the laser through the liquid filled HOF and observed the ejected liquid droplet patterns using a CCD camera by varying major experimental parameters: the laser power (P), irradiation time (T), and the vertical distance (Z) between the HOF and substrate. The power and irradiation time of the laser were controlled by Labview in a personal computer, and the vertical distance was manually controlled via a xyz translational stage.

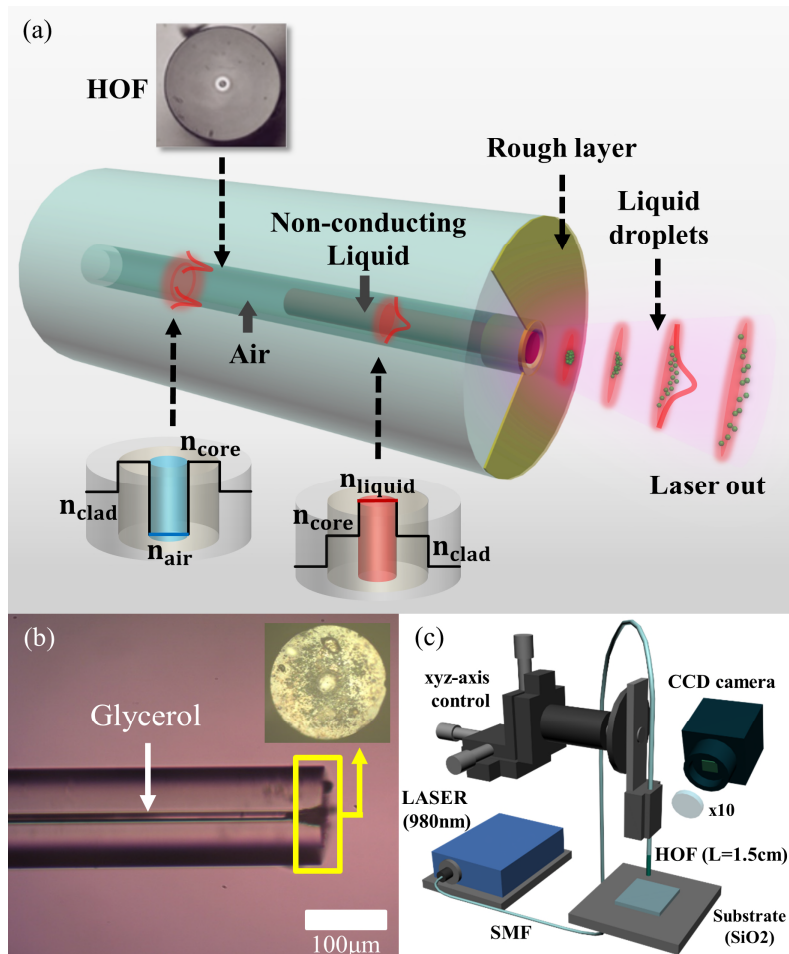

Fig. 1. (a) Schematic diagram of laser-driven jetting of nanoscale non-conducting liquid droplets via hollow optical fiber (HOF). (b) Micrograph of the HOF filled with the glycerol. Inset shows the image of the end facet of the HOF where diamond nanoparticles have been attached. (c) Schematic diagram of the experimental set-up. Laser (λ=980nm) was launched through single mode fiber (SMF) and it was spliced to where HOF has been spliced. The laser power and irradiation time have been controlled by Labview, while the distance between HOF and substrate manually using xyz translational stage.

Fig. 2(a) shows an experimental observation for the liquid droplets ejected from the HOF and their distribution on the silica substrate. The laser power (P) was 22.1±0.5mW with irradiation time (T) of 1s and vertical distance between the HOF and substrate (Z) of 50μm. The gray scale diagram across the central horizontal line is shown in Fig. 2(b) and it agrees well with a Gaussian fitting shown in a red curve. In a prior report on optical levitation of liquid droplets [23], Gaussian-like distribution in the droplets has been also reported such that levitated droplet size decreased as the observation position moved toward the edge of light intensity distribution [23]. As shown in Fig. 2, our atomized droplets carried by the light were similarly found to follow the laser beam profile out of the HOF such that the droplet size near the center where the light intensity was higher is significantly larger than those near the edges where the light intensity was week.

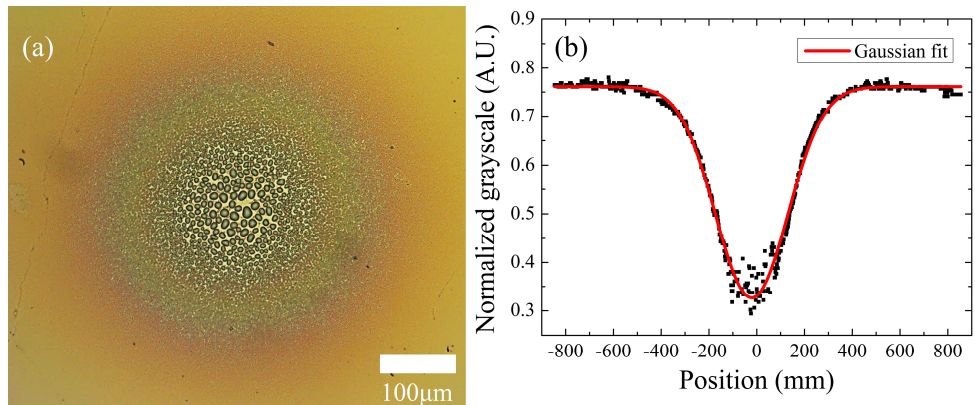

Fig. 2. (a) Optical micrograph of glycerol droplets optically ejected from HOF and deposited on a SiO$_2$ substrate. P=22.1±0.5mW, T=1s and Z=50μm (color online). (b) The normalized grayscale of Fig. 2(a) across the central horizontal line, and its Gaussian fit shown in the red curve.

In order to further investigate the characteristics of our unique laser-driven jetting, we captured the liquid ejection motions using a CCD camera (50fps). Captured-images are sequentially arranged in Fig. 3. Here, we set P=20±2mW, T=1s, and Z=400μm. We observed that the liquid droplets were propagating in a dark grey belt distributed in a spherical cone shape. Note that this is very similar to the equi-phase front of the light wave from the optical fiber, which defines its numerical aperture [24]. The close overlap between the droplet distribution and light propagation patterns strongly indicates that the light momentum within and out of HOF plays a key role in this unique all-laser driven atomization and spraying.

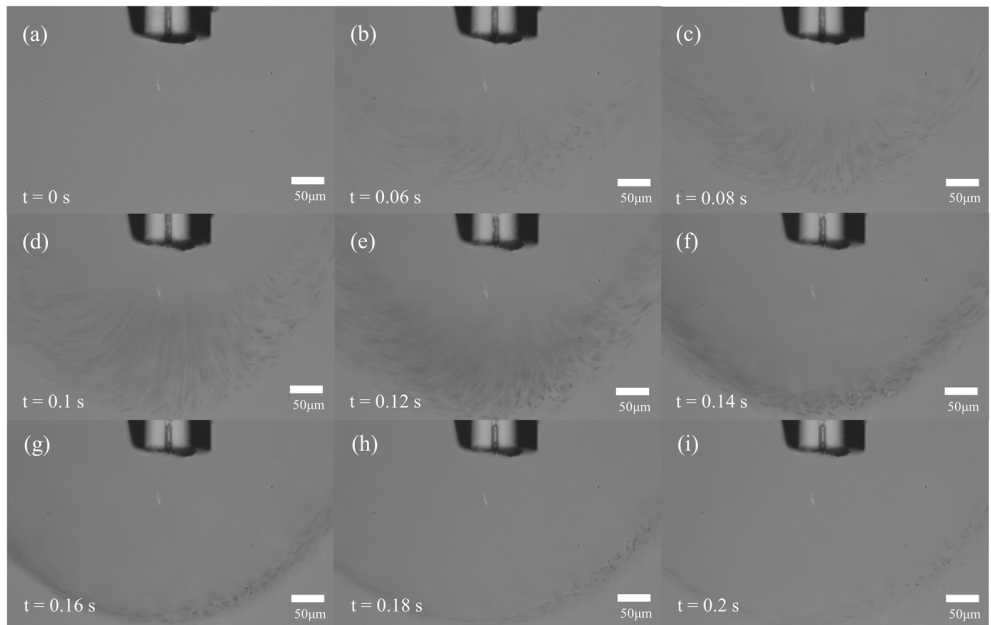

Fig. 3. Frames of the laser-driven liquid droplet ejection process. (a) t=0 s, (b) t=0.06s, (c) t=0.08s, (d) t=0.1s, (e) t=0.12s, (f) t=0.14s, (g) t=0.16s, (h) t=0.18s, (i) t=0.2s. The droplets were observed as a dark gray belt distributed in a spherical cone shape.

We then experimentally investigated the impacts of the laser on the distribution of the liquid droplets on SiO$_2$ substrate as a function of the vertical distance between the HOF and substrate, and the normalized results are shown in Fig. 4(a). Here, we set P=11±2mW and T=0.2s. We defined 4-sigma, four times of standard deviation of Gaussian fitting, shown in Fig. 2(b), 95% of Gaussian fitting area as the diameter of the liquid distributed on the silica substrate. The diameter of the ejected droplet distribution area increased proportional to Z, the vertical distance between HOF and the substrate, as summarized in Fig. 4(a).

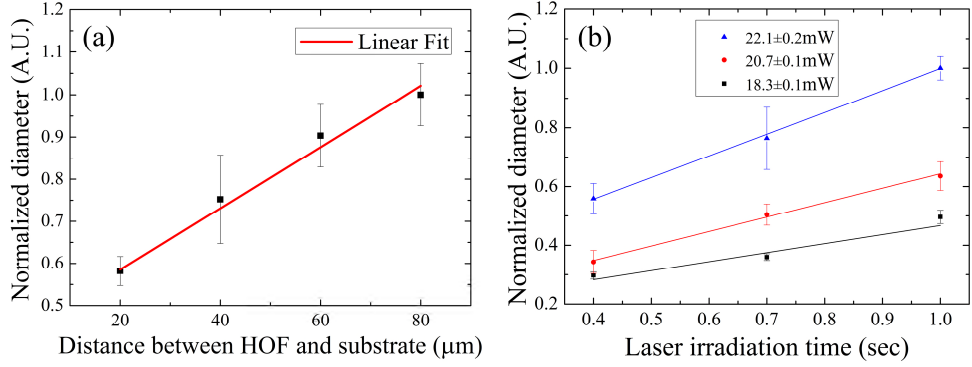

Fig. 4. The influence of the laser on the distribution of the liquid droplets. (a) The normalized diameter of the liquid droplet distributed area as a function of the vertical distance between the HOF and SiO$_2$ substrate with P=11±2mW and T=0.2s. (b) The normalized diameter of the liquid droplet distributed area as a function of the irradiation time for various laser powers with Z=50μm. Inset shows the various optical powers, and the data have been linearly fitted.

For an optical fiber with a given numerical aperture (N.A.), the light going out of the core is confined into a cone, whose angle satisfies [25]: $N.A. = n\sin\theta = \sqrt{n_{core}^2 - n_{clad}^2}$. Here $n$ is the refractive index of the medium light is propagating, $n_{core}$ and $n_{clad}$ are the refractive indices of the core and cladding of the optical fiber, respectively. The light propagates within this spherical cone with the maximum angle of $\theta$, and the laser beam diameter increases linearly with the axial distance, which explains the linear behavior in Fig. 4(a). The figure shows a close correlation between the light propagation and the droplet distributed area, which strongly indicates that the light from HOF plays a key role in transporting the droplets. We have also measured the impact of the laser irradiation time on the distribution of the liquid droplets for a fixed axial distance of Z=50μm, and the normalized results are summarized in Fig. 4(b) for various laser powers. It was observed that the diameter is linearly increasing proportional to the irradiation time for all laser powers.

As is well known, the radiant exposure of the laser, $H = \int_0^T E_e dt$ where $E_e$ is irradiance, the flux of radiant energy per unit area, and T is the irradiation time, linearly increases with the irradiation time for CW laser [26]. Due to this effect with tightly focusing the laser into the liquid, we expect the light energy would transfer its momentum into the liquid making the linear relationship between the droplet distribution and the irradiation time, as can be seen in Fig. 4(b). We also observed that this jetting process requires a certain level of laser power. Note that the glycerol itself has a very little absorption [16], and the liquid length within is HOF only ~300μm, therefore the material absorption and consequent heat generation could be negligible. Furthermore, the level of laser power is moderately low enough and the irradiation time is less than a second, which will not generate nonlinear optic effects. The exact mechanism of this laser power dependence is not clearly understood yet, and it is being investigated by the authors.

We now focus on individual droplet size within the Gaussian-like distributed area. To measure the size of each individual droplet, we mixed gold nanoparticles in glycerol to make a colloidal suspension. Here, the gold nanoparticles we used are gold nanorods (GNRs) and it was prepared by previous methods [27]. Briefly, the seed solution was prepared by mixing cetyltrimethylammonium bromide (CTAB, 0.2M, 5mL) and $HAuCl_4$ (0.5mM, 5mL) with freshly prepared ice-cold $NaBH_4$ (10mM, 600mL). The seed solution was then used for the synthesis of GNRs 3~5h after preparation. CTAB (0.2M, 20.0mL) was mixed with silver nitrate (10mM, 120mL) and $HAuCl_4$ (1mM, 20.0mL) in two separate flasks and gently mixed and ascorbic acid (0.10M, 240mL) was added. To this mixture, the seed solution (48mL) was finally added to initiate growth to yield GNRs with an aspect of approximately 6.1. The GNRs were average length and width of 55±5nm and 9±1nm, respectively. A suspension of GNRs containing CTAB was maintained at a constant temperature at 37˚C. GNRs in suspension were centrifuged at 15,000rpm for 10mins, and concentrated to remove excess CTAB for three times. And then GNRs were re-suspended in glycerol. It has been recently reported that gold nanoparticles do not react with glycerol chemically, and the colloid maintains inherent physical properties of both gold nanoparticles and glycerol [28]. With the gold nanoparticle-loaded glycerol, we repeated the same laser jetting experiment to obtain droplet distribution similar to Fig. 2, and dried it for 4 hours in an oven at 25˚C.

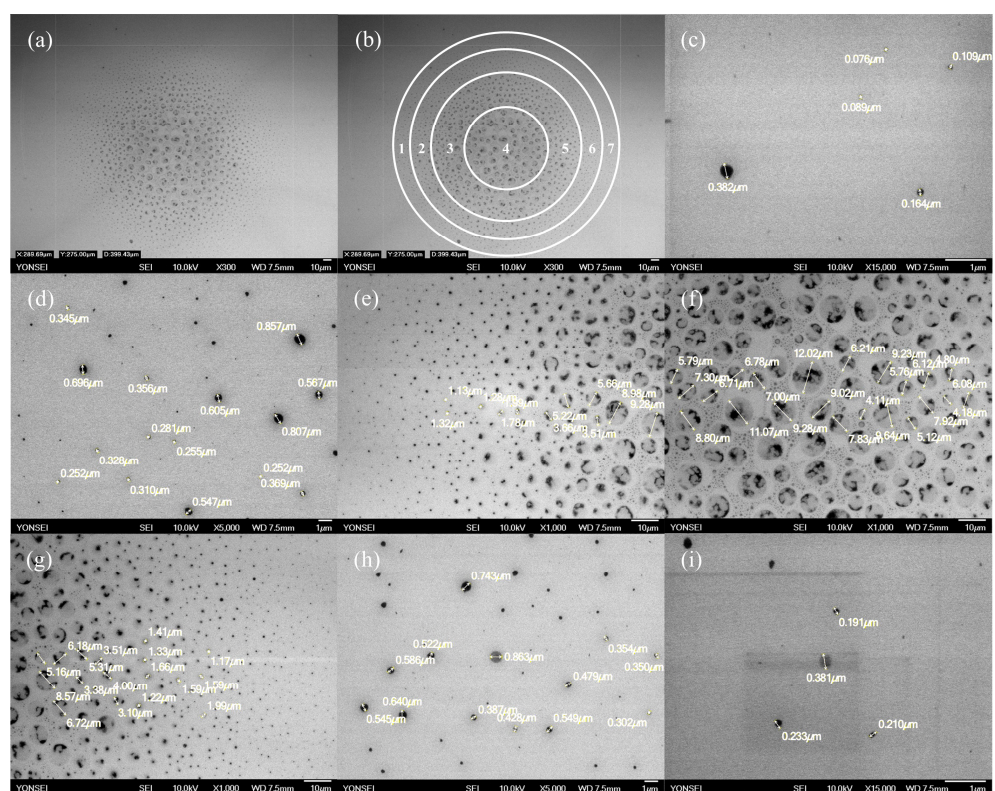

Fig. 5. (a) Scanning electron microscope (SEM) micrograph of gold nanorods that were contained in individual glycerol droplets. (total scanned area: 289.69 x 275μm$^2$) (b) Seven divided sections (1~7) of Fig. 5(a) for each droplet size measurement. (c) Section 1, (d) Section 2, (e) Section 3, (f) Section 4 (g) Section 5, (h) Section 6, (i) Section 7 of Fig. 5(b).

The gold nanoparticles contained in each droplet can represent its dimension as the glycerol evaporates, which is well-explained by 'coffee-ring effect' [29]. The gold nanoparticle patterns in individual droplets were measured by scanning electron microscope (SEM) and the results are summarized in Fig. 5. Here, we set P=189.7±0.7mW, T=0.1s and Z=150µm. Gold nanoparticle patterns in Fig. 5(a) were similar to the glycerol droplet pattern in Fig. 2(a), as expected. We divided Fig. 5(a) into seven sections as in Fig. 5(b), section 1 to 7. The individual droplet size, represented by gold nanoparticles, in each section is shown in Fig. 5(c)~(i). In the sections near the edge, the droplet diameter ranged from 76 to 400nm and it increased to 1 to 11µm near the central section 4. We experimentally confirmed that our laser jetting technique can provide a wide range of liquid droplet diameter from a few tens of nm to a few µm, in a highly localized area less than 300 x 300µm$^2$.

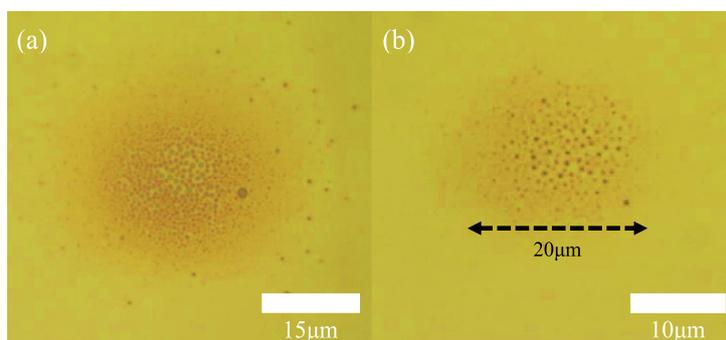

Fig. 6. (a) Virgin olive oil droplet distribution on SiO$_2$ substrate (b) Highly localized virgin olive oil droplet distribution within the diameter of about 20µm.

After confirming unique capability of non-conducting liquid droplet jetting using glycerol, a kind of polyol, we further assessed the feasibility of jetting larger hydrophobic molecules than glycerol, olive oil (fat or lipids) on SiO$_2$ substrate. As shown in Fig. 6(a) we could successfully obtain laser-driven jetting of olive oil. Note that prior jetting techniques, including EHD have not been able to accommodate this liquid. This can be applied to bio-chemical applications for highly hydrophobic compounds of high molecular weight [30]. Furthermore, this technology can have strong potentials in various applications that require jetting or printing of non-conducting liquids, such as particle and drug delivery in microscopic environment without electrodes [31, 32].

Also, we have experimentally investigated the minimum droplet distribution area achievable using our laser jetting technique with virgin olive oil about 20µm, shown in Fig. 6(b). This strongly indicates that our laser-driven jetting can achieve a higher resolution of patterning than conventional ink jet printing based on piezoelectric or thermal actuation. Drop-on-demand style EHD jetting has a higher patterning resolution of a few hundred nanometer range. Since nanometer scale droplets can be generated by our laser jetting technique, finer control of optical jetting parameters could provide the similar level of resolution.

## 3. Conclusion

In summary, we experimentally demonstrated laser-driven jetting of non-conducting transparent liquid (glycerol and olive oil) out of surface-treated hollow optical fiber (HOF), producing liquid droplets ranging from 76nm to 11μm. These droplets were carried by the propagating light field forming a spherical cone, which were then deposited on a silica substrate in a Gaussian distribution. We found that successful jetting required additional preparation such that the end facet of the HOF and its inside surface should be covered with a rough surface layer, which was achieved by depositing nano-diamond on the surfaces. The liquid volume used for the laser-driven jetting was in the range of ~2.6 pico-liter, which was fed to HOF by the capillary force. The deposited area diameter has shown a linear dependence on the process parameters such as the laser irradiation time and the vertical distance between HOF and the substrate. The droplet deposited area diameter was flexibly controlled with minimum droplet distribution of 15μm, which makes this unique all-optical fiber-based jetting technology versatile and flexible. Moreover, this system can precisely generate the femto-scale atomized liquid droplets. This technique can obviate imperative requirement of electrodes as in prior electric hydrodynamic printing and can be readily implemented in-situ microscopic environment.